\documentclass{ISMA_USD}

\usepackage{subcaption}
\usepackage{nicefrac}
\usepackage{booktabs}
\usepackage{tikz}
\usepackage{algorithm}
\usepackage{algpseudocode}
\usetikzlibrary{bayesnet,positioning,arrows.meta}

\newcommand{\B}[1]{\bm{#1}}

\newcommand{\tran}{^{\mkern-1.5mu\mathsf{T}}}

\graphicspath{{./figures}}

\hypersetup{
  pdftitle  = {Latent variable models for simultaneous EOV identification and removal in population-based SHM},
  pdfauthor = {M. D. Champneys, M. R. Jones, A. J. Hughes, T. J. Rogers, E.J. Cross, K. Worden},
  pdfkeywords = {PBSHM, EOV, Gaussian processes, latent variable models, damage detection, offshore wind}
}

\title{Latent variable models for simultaneous EOV \\identification and removal in population-based SHM}

\author[1,2]{M. D. Champneys}
\author[1]{M. R. Jones}
\author[1]{A. J. Hughes}
\author[1]{T. J. Rogers}
\author[1]{E. J. Cross}
\author[1]{K. Worden}

\affil[1]{Dynamics Research Group, University of Sheffield, Mappin Street, Sheffield S1 3JD, UK}
\affil[2]{Centre for Machine Intelligence, University of Sheffield, Mappin Street, Sheffield S1 3JD, UK}

\date{}

\begin{document}

\abstract{
  The robust treatment of environmental and operational variability (EOV) is an open challenge in population-based structural health monitoring (PBSHM). The difficulty is compounded in the case that the EOV signals are unmeasured. A common approach in conventional SHM is to apply \emph{projection-based} methods that discard subspaces of healthy feature data, reasoning that the EOV signal dominates the variance of the measured features. However, a common pitfall of projection-based approaches is that when damage acts close to the same variance-dominant direction, damage sensitivity is removed along with the EOV. An alternative identifying assumption for the removal of particular unmeasured EOVs is \emph{slowness}; the latent EOV process is characterised by its long temporal correlation. In this paper, the latent EOV is cast as a state-space Gaussian process, enabling tractable $\mathcal{O}(T)$ inference via a Kalman filter. A robust hierarchical Bayesian identification framework is developed that enables population-level identification of latent EOVs and EOV-free residual features, using a Laplace approximation. The approach is first validated on a single laboratory-scale benchmark structure from the literature, subject to thermal EOVs, demonstrating robust damage detection and EOV recovery. The method is then applied to a simulated nine-turbine offshore wind farm with staggered deployment and damage, where it delivers a substantial true-positive uplift over projection and cointegration-based baselines at matched false-positive rates.
}

\maketitle

\section{Introduction}

Structural health monitoring (SHM) seeks to detect, locate, and characterise structural damage from measured response data \cite{farrar2012structural}.
A persistent obstacle is \emph{environmental and operational variability} (EOV); ambient temperature, wind speed, and operational load shifts all alter the measured feature distribution, causing false alarms if damage detection is framed as distributional outlier detection on the raw features \cite{sohn2002statistical}. 

To date, many approaches have been proposed to manage the effects of EOVs \cite{sohn2007effects}. The central distinction between existing paradigms is the observability of the EOV. In the case that the EOV signal is observed (e.g.\ measured temperatures, operational logs etc.), \emph{regression}\footnote{Also called explicit methods by some authors.} approaches model the feature-EOV relationship directly and detect damage in the regression residuals \cite{peeters2001z24}. 

When the EOV is unmeasured\footnote{Also called implicit methods by some authors.}, the EOV subspace must be inferred from the feature data itself. The dominant paradigms are cointegration and projection-based removal. In the former case, the existence of a \emph{cointegrating relationship} is assumed, whereby the features are individually non-stationary, driven by a shared integrated environmental trend, yet a linear combination of them is stationary \cite{cross2011cointegration, cross2012cointegration}. Damage is inferred by the violation of stationarity. Latent-variable treatments have also been proposed; factor analysis has been used to eliminate unmeasured EOVs and to distinguish their effects from those of sensor faults and damage \cite{kullaa2009missing, kullaa2011distinguishing}.

This paper focuses on the case of unmeasured EOVs. For projection-based approaches such as \emph{minor component analysis} (MCA) \cite{yan2005partI, yan2005partII}, the identifying assumption is \emph{variance-dominance}, the top principal components (or other similarly-identified subspaces) of healthy feature data are discarded, with the reasoning that they are dominated by the EOV. However, for structural systems where damage induces proportional shifts across all monitored vibration modes (e.g.\ global foundation degradation from scour or corrosion) the damage direction may be close to collinear with the EOV direction, and the removal of the high-variance subspace removes the damage signature alongside the EOV.

An alternative to variance-dominance is \emph{slowness}; the identifying assumption is that the EOV varies on longer timescales than damage, noise or the structural dynamics, so it can be identified by a latent process with a long-correlation prior, regardless of its variance signature \cite{wiskott2002slow, turner2007maximum}. So-called slow feature analysis (SFA) is an approach also used for nuisance-variable removal in other monitoring domains; e.g.\ condition monitoring \cite{zhang2019sfa}. A common limitation of slowness-based removal, however, is that abnormalities with a gradual onset (e.g.\ crack onset and propagation) can themselves be absorbed by the slow latent process, masking the presence of damage.

The challenges facing both variance-dominance and slowness approaches can be amplified in a population-based setting. Structures in a population may share a common environmental driver yet differ in their per-structure response to EOVs. Furthermore, assets commissioned (or instrumented) at different times accumulate healthy training data at different rates, leaving newly deployed members with poorly estimated loadings at the critical early-detection stage. However, methodologies that share data between structures can improve robustness to EOVs compared to the single-structure setting.

\paragraph{Contribution}

This work proposes a shared Gaussian-process latent EOV (GLEOV)\footnote{Pronounced (with some difficulty) as `glove'.} model for simultaneous identification and removal of EOVs across a population of structures. The latent EOV signal is modelled as a Mat\'{e}rn Gaussian process (GP) with a long lengthscale. The result of \cite{hartikainen2010} is used to cast the GP in a linear-Gaussian, state-space form, enabling efficient computation of the marginal likelihood and model predictions via a Kalman-filtering framework. The same state-space treatment of Gaussian processes underpins the Gaussian-process latent force models used for joint input-state-parameter estimation in structural dynamics \cite{rogers2020application}.

In the proposed approach, the latent EOV is identified by its temporal correlation structure rather than its variance, avoiding the limitations of variance-dominance as an assumption for EOV removal. The latent EOV is shared across the population under a hierarchical prior on per-structure EOV loadings. Data-rich structures anchor the shared EOV and lend statistical strength to data-poor neighbours. Damage is detected using the Kalman innovations (the residuals on the projected EOVs) as damage sensitive features, with an approximate Laplace posterior supplying an uncertainty-aware exceedance probability at any given threshold.

The principal contributions of this work are:

\begin{itemize}
  \item A Gaussian-process latent EOV (GLEOV) model for efficient simultaneous identification and removal of EOVs in SHM.
  \item A hierarchical Bayesian extension to the case of populations of structures.  
  \item Validation on an experimental single-structure benchmark from the literature, and evaluation on a nine-turbine synthetic offshore wind farm with staggered deployment and damage acting along the EOV direction, demonstrating true-positive uplift over projection-based baselines at matched false-positive rates.
\end{itemize}

\paragraph{Related work}
  
The problem of accounting for EOVs in SHM has been well studied in the literature \cite{worden2020brief}. Much of the research effort has been focussed on the treatment of single structures and observed EOVs \cite{avendano2020gaussian, worden2018switching, moller2026greybox}. However, there are several works that consider unobserved EOVs. In \cite{cross2012features}, a projection-based approach based on minor-component analysis is compared to other implicit methods including cointegration \cite{cross2011cointegration, cross2012cointegration}. Although successful, these results are limited to a single structure and do not recover the latent EOV signal. 

Several approaches based on SFA (that can recover unobserved and observed EOVs) have also been proposed in process monitoring and SHM \cite{zhang2019sfa, wang2026dynamic}, although these also have been focussed on single structures. 

Additional single-structure treatments are to be found in the literature, using similar identifying assumptions. A localised principal-component method for environment-induced bridge modal variability \cite{Zhen2022sfapca} uses a variance-dominance argument and applies the projection locally rather than globally, while a nonlinear system-identification approach for output-only monitoring under changing environments \cite{reynders2014output} removes the environmental influence implicitly via kernel PCA, a nonlinear extension of the variance-dominance assumption. As with the works above, both are confined to a single structure and share no information across a population. A Kalman-filtering approach is adopted in \cite{erazo2019vibration}, where the filter innovations serve as damage-sensitive features under changing environmental conditions, although again for a single structure and without recovery of the latent environmental signal. In \cite{aravanis2020stochastic}, the unmeasured EOV instead enters a random-coefficient functional model as a latent scheduling variable, although it is marginalised rather than recovered. What unites all of these methods is that the environmental process is treated as a nuisance to be removed. Conversely, the authors of \cite{zhao2026disentangling} invert this view, treating the slow latent process as the signal rather than the nuisance and inferring a monotonic degradation state from fast operational dynamics with a hierarchical neural controlled differential equation.

The treatment of EOVs in a population-based context has had comparatively less attention in the literature. The foundations of PBSHM are set out in \cite{bull2021foundations1, gosliga2021foundations2, gardner2021foundations3}, and in \cite{worden2020brief} the major challenges are outlined, including the treatment of EOVs. An earlier population treatment is given in \cite{vamvoudakis2018population}, using unsupervised multiple-model time-series methods across nominally-identical structures, but without an explicit EOV model. Several papers employ hierarchical Bayesian methods to population-based SHM \cite{dardeno2023population, DARDENO2024111072}. While these approaches are able to recover functional relationships between observed features and EOVs, they rely on at least partial observation of the EOV signal, in contrast to the fully-unobserved setting considered here. Closer to the present setting, a hierarchical Bayesian approach to anomaly detection in the natural frequencies of offshore wind turbines under temperature variation is presented in \cite{smith2024anomaly}, although the EOV is not there recovered as a latent signal.

Two relevant works in this vein are a GP regression-surface approach to EOVs in \cite{lin2020towards, lin2022mapping} and an adaptive methodology in \cite{qu2025multi} that combines data from two nominally-identical structures, governed by a mixing hyperparameter, to build adaptive stochastic models of the damage-sensitive features. This latter work demonstrates results on a population of two structures in a manner related to the present work, but it does not recover the underlying latent EOV signal.

Perhaps closest to the present work is the single-structure method of \cite{zhu2022structural}, which combines a Gaussian process with probabilistic PCA to account for both observed and latent EOVs. There, the GP models the structural response as a function of the \emph{measured} EOVs, while PCA removes the \emph{unmeasured} component. By contrast, GLEOV represents the unobserved EOV itself as a slow latent Gaussian process that requires no EOV observations, shares this latent process across a population of structures under a hierarchical prior, and retains $\mathcal{O}(T)$ state-space inference.

Although many approaches have been proposed to address the problem of EOVs in SHM, the combination of slow-latent EOV removal with population-level information sharing, so that data-rich structures anchor the shared EOV and provide an informative prior for data-poor neighbours, has not been treated in a unified probabilistic framework.
  
\paragraph{Paper structure} The remainder of this paper is structured as follows: The following section describes the proposed method. A third section presents two case-study examples of the approach on a single-structure laboratory dataset and a simulated population of structures respectively. A final section presents some discussion and conclusions.

\section{The Gaussian-process latent EOV (GLEOV) model}

\paragraph{Generative model}

In this work, a data-generating process of the following form is assumed. For structure $i \in \{1,\ldots,N\}$ at time $t$, the observed $M$-dimensional feature vector $\B{x}_{i,t} \in \mathbb{R}^M$ is modelled as,

\begin{equation}
  \B{x}_{i,t} = \B{\mu}_i + W_i (\B{z}_t + \B{\rho}_{i,t}) + \B{\epsilon}_{i,t},
  \quad
  \B{\epsilon}_{i,t} \sim \mathcal{N}\left(\B{0},\; \sigma_e^2 \B{I}_M \right),
  \quad 
  \B{\rho}_{i,t} \sim \mathcal{N}\left(\B{0},\; \tau_T^2 \B{I}_K\right),
  \label{eq:obs}
\end{equation}

where $\B{\mu}_i \in \mathbb{R}^M$ is the per-structure mean, $\B{z}_t \in \mathbb{R}^K$ are the shared slow latent EOVs. Intuitively, this treats the observed features in time as a Gaussian, i.i.d.\ process, per-structure, corrupted additively by a correlated latent process $\B{z}$ via the action of the per-structure EOV loading weights $W_i$. The $\B{\rho}_{i,t}$ term accounts for the fact that even shared EOV signals (e.g.\ temperature) may vary stochastically across the population.  

Note that the above can be re-written in the form,

\begin{equation}
  \B{x}_{i,t} = \B{\mu}_i + W_i \B{z}_t + \B{\nu}_{i,t},
  \quad
  \B{\nu}_{i,t} \sim \mathcal{N}\!\left(
    \B{0},\; \sigma_e^2 \B{I}_M + \tau_T^2 W_i W_i\tran
  \right),
  \label{eq:obs2}
\end{equation}

\begin{figure}[ht]
  \centering
  \begin{tikzpicture}

    \node[obs](x){$\B{x}_{i,t}$};
    \node[latent, left=of x](z){$\B{z}_t$};
    \node[latent, above=of x](mui){$\B{\mu}_i$};
    \node[latent, right=of mui](Wi){$W_i$};
    \node[latent, right=of Wi](W0){$W_0$};
    \node[latent, below=0.5cm of W0](sig_e){$\sigma^2_e$};
    \node[latent, below=0.5cm of sig_e](taut){$\tau^2_T$};

    \draw[->] (z.70) to[out=70,in=110,looseness=10] (z.110);

    \edge {Wi,mui,z, sig_e, taut} {x};
    \edge {W0} {Wi};
    \plate {pk} {(z)} {$K$};
    \plate {pt} {(z)(x)} {$T$};
    \plate {pn} {(Wi)(mui)(x)} {$N$};

  \end{tikzpicture}
  \caption{Probabilistic graphical model of the GLEOV generative process. The
  shared latent EOV $\B{z}_t$ evolves as a temporal Gaussian process (self-loop)
  and is mapped to the observed features $\B{x}_{i,t}$ (shaded node) by the per-structure
  loading $W_i$, offset by the per-structure mean $\B{\mu}_i$; loadings are drawn
  from a population consensus $W_0$. Plates index the latent dimension ($K$),
  time ($T$), and structure ($N$).}
  \label{fig:pgm}
\end{figure}
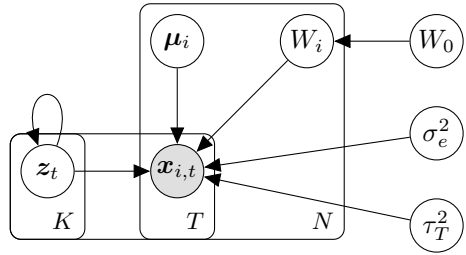

Thus, given estimates of the weightings, means and latent EOV signal, an EOV-removed residual can be computed as,

\begin{equation}
  \B{\nu}_{i,t}
  = \B{x}_{i,t} - \hat{\B{\mu}}_i - \hat{W}_i \hat{\B{z}}_{t \mid t-1},
  \label{eqn:innova}
\end{equation}

where $\hat{\cdot}$ denotes an estimated quantity. In the absence of damage, this residual is Gaussian and independent across time, with an approximately stationary covariance, making it an ideal candidate for novelty detection in SHM. A probabilistic graphical model of the proposed data-generating process is given in Figure \ref{fig:pgm}.

\paragraph{Priors} 

In this work, the shared latent $z_t$ is given a zero-mean Matérn-3/2 GP prior\footnote{In the exposition of this section, scalar $z_t$ is considered only. However, the approach can be easily extended to vector-valued GPs by inflating the size of the state space.}. It can be shown that such a GP can be written as a linear-Gaussian state-space-model as \cite{hartikainen2010},

\begin{equation}
  \begin{bmatrix}z_t\\ \dot{ z}_t \end{bmatrix} = 
  A_d \begin{bmatrix} z_{t-1}\\ \dot{ z}_{t-1} \end{bmatrix} + \mathbf{w}_t,
  \qquad
  \mathbf{w}_t\sim\mathcal N(\mathbf 0, Q_d),
  \qquad 
  \begin{bmatrix} z_0\\ \dot{ z}_0 \end{bmatrix}\sim\mathcal N(\mathbf 0, P_\infty)
\end{equation}

\begin{equation}
  A_d = e^{-\lambda \Delta_t}\begin{bmatrix}1+\lambda\Delta_t & \Delta_t\\ -\lambda^2\Delta_t & 1-\lambda\Delta_t\end{bmatrix},
  \quad  
  P_\infty=\begin{bmatrix}1 & 0 \\ 0&\lambda^2 \end{bmatrix},
  \quad 
  Q_d= P_\infty- A_d P_\infty A_d\tran
  \label{eqn:ssm}
\end{equation}

Here $\lambda=\sqrt3/\ell$, where $\ell$ is the GP lengthscale and $\Delta_t$ the sampling period. The matrix $P_\infty$ is the stationary covariance of the state at which the filter is initialised so that the recursion begins from the GP's stationary distribution. Note that in order to avoid identifiability issues with the magnitude of the loading vectors, the signal variance of the GP is here assumed to be equal to 1 without loss of generality.

The loadings receive an informative hierarchical prior, $W_i \sim \mathcal{N}(W_0,\, 10^{-6}\,\B{I})$, it is reasoned here that the EOV loading direction $W_i$ will be highly similar between nominally identical structures, and that data-poor structures must borrow information from the population in order to identify it accurately. In contrast, the per-structure means receive a deliberately uninformative prior, $\B{\mu}_i \sim \mathcal{N}(\B{0},\, 10^{2}\,\B{I})$, with the reasoning that the mean values will be structure specific and not necessarily shared among population members. All prior distributions and hyperparameters are listed in Table \ref{tbl:priors}.

Both the GP lengthscale and time step are considered fixed in this work; the lengthscale is expressed in units of the sampling period $\Delta_t$. In practice, the value of the lengthscale can be set according to knowledge of the population. For example, an annual seasonal trend could be given a lengthscale parameter in the range 50--150 days, whereas daily variation would require a shorter lengthscale on the order of hours. Note that the fixed value of the lengthscale encodes the assumption that the dynamics of the latent GP are slow. The robustness of the method to this choice is confirmed by a sensitivity study presented in Appendix~\ref{app:lengthscale}.

\begin{table}[h]
  \centering
  \caption{Prior distributions and fixed hyperparameters of the GLEOV model.}
  \label{tbl:priors}
  \begin{tabular}{l|ll}
    \hline
    Parameter & Symbol & Prior / value \\\\
    \hline
    \multicolumn{3}{l}{\emph{Priors}} \\
    \hline
    \quad Per-structure mean            & $\B{\mu}_i$ & $\mathcal{N}(\B{0},\, 10^2 \B{I})$ \\
    \quad Per-structure loading         & $W_i$       & $\mathcal{N}(W_0,\, 10^{-6} \B{I})$ \\
    \quad Consensus loading             & $W_0$       & flat ($\propto 1$) \\
    \quad Measurement-noise scale       & $\sigma_e$  & flat ($\propto 1$) on $\log\sigma_e$ \\
    \quad Population EOV scale           & $\tau_T$    & $\log\tau_T \sim \mathcal{N}(\log 0.1,\, 1)$ \\
    \hline
    \multicolumn{3}{l}{\emph{Hyperparameters (fixed)}} \\
    \hline
    \quad GP lengthscale                & $\ell$      & $100$ (units of $\Delta_t$) \\
    \quad Sampling period               & $\Delta_t$  & $1$  \\
    \quad Latent EOV dimension          & $K$         & $1$ \\
    \quad Gating false-positive rate    & $\alpha_\text{gate}$ & $0.01$ \\
    \quad Detection false-positive rate & $\alpha$    & $0.001$ \\
    \hline
  \end{tabular}
\end{table}

\subsection{Inference}

The model is learned from data using a recursive Bayesian estimation approach based on a Kalman filter. Because the GP prior is discretised exactly as a linear-Gaussian SSM, $z_t$ can be marginalised  (conditional on the parameters $\theta=\{\mu, W, W_0, \sigma_e, \tau_T\}$) in closed form \cite{sarkka2023bayesian}. Let $m_t$ and $P_t$ be the filtering mean and covariance at time $t$. Let also $H$ be a suitable selection matrix such that $z_t = H m_t$. Then the Kalman filter equations are given by,

\begin{equation}
  p\!\left(\begin{bmatrix} z_t \\ \dot{z}_t \end{bmatrix} \,\middle|\, x_{1:t-1}, \theta\right)
    = \mathcal{N}(m^-, P^-),
  \qquad
  m^- = A_d m_{t-1},
  \qquad
  P^- = A_d P_{t-1} A_d^\top + Q_d.
\end{equation}

\begin{equation}
  p(x_t \mid x_{1:t-1}, \theta) = \mathcal{N}(\mu + W \hat{z}_{t \mid t-1}, \; S_t),
  \qquad
  \hat{z}_{t \mid t-1} = H m^-,
  \qquad
  S_t = W H P^- H^\top W^\top + R
  \label{eqn:St}
\end{equation}

where, 

\begin{equation}
  R = \operatorname{blockdiag}(R_1, \ldots, R_N), \qquad R_i = \sigma_e^2 \B{I}_M + \tau_T^2 W_i W_i^\top
\end{equation}

is the block-diagonal observation noise model for each structure in the population.

\begin{equation}
  p\!\left(\begin{bmatrix} z_t \\ \dot{z}_t \end{bmatrix} \,\middle|\, x_{1:t}\right)
    = \mathcal{N}(m_t, P_t),
  \qquad
  m_t = m^- + K_t \nu_t,
  \qquad
  P_t = P^- - K_t S_t K_t^\top,
\end{equation}

with Kalman gain $K_t = P^- H^\top W^\top S_t^{-1}$ and innovation $\nu_t = x_t - \mu - W \hat{z}_{t \mid t-1}$. The marginal likelihood can thus be computed recursively as the product of the one-step-ahead predictive densities,

\begin{equation}
  p(x_{1:T} \mid \theta) = \prod_{t=1}^{T} \mathcal{N}(x_t \mid \mu + W \hat{z}_{t \mid t-1}, \; S_t).
\end{equation}

However, the above is still conditional on the unknown parameters $\theta$. Unfortunately, the full posterior distribution is intractable, and so a Laplace approximation is employed.   

\paragraph{MAP} The maximum \emph{a posteriori} (MAP) values of the unknown parameters $\hat{\theta}$, are computed by minimisation of the negative log joint distribution,

\begin{equation}
\hat{\theta} = \arg\min_{\theta} - \log p(x_{1:T} \mid \theta) - \log p(\theta) %\propto p(\theta | x_{1:T})
\label{eqn:map}
\end{equation}

where $p(\theta)$ is given by the product of the parameter priors above. Here, the optimisation is run using the L-BFGS-B solver \cite{byrd1995lbfgsb, zhu1997lbfgsb}, as implemented in \texttt{scipy} \cite{2020SciPyNMeth}. 

To aid convergence, the optimisation is preconditioned: L-BFGS-B is run in a whitened space in which each parameter block is rescaled by a counting-heuristic estimate of its Fisher information ($\propto 1/\sqrt{n_\text{obs}}$). Without this step the gradient magnitudes across the heterogeneous blocks (loadings, means, and log-scaled noise terms) differ by orders of magnitude and the optimiser fails to converge.

The optimiser is initialised at the pooled-PCA solution: The per-structure means $\B{\mu}_i$ are set to the observed training-window means, and the consensus loading $W_0$ and every per-structure loading $W_i$ are initialised to the top-$K$ right singular vectors of the per-structure mean-centred training features, pooled across all structures. The noise scales are initialised to a fraction of the residual spread ($\sigma_e$ at one tenth of the pooled standard deviation) and to $\tau_T = 0.1$. From this warm start, L-BFGS-B refines the loadings under the GP prior, replacing the purely variance-based initial subspace with one that additionally respects the temporal-correlation structure of the latent. Because each marginal-likelihood evaluation is linear in the series length, the fit is inexpensive in practice; the full nine-turbine population model here (Section~\ref{sec:owt}, $365$ observations) converges in approximately $70$\,s on a single CPU core.

\paragraph{Laplace approximation}

The Laplace approximation is a Gaussian surrogate centred at the MAP estimate, with covariance the inverse Hessian of the negative log-joint at the mode,

\begin{equation}
  p(\theta \mid \B{x}_{1:T}) \;\approx\; \mathcal{N}\!\big(\hat{\theta},\,\hat\Sigma\big),
  \qquad
  \hat \Sigma \;=\; \big[-\nabla^2_\theta \log p(\theta,{x}_{1:T})\big|_{\hat\theta}\big]^{-1}
  \label{eqn:laplace}
\end{equation}

The Hessian matrix is evaluated exactly by automatic differentiation of the negative log-joint at the mode, and hence $\hat \Sigma$ is obtained by its inversion.

\subsection{Damage detection}
 
Once the model is identified, the EOV-removed residual for each structure is the Kalman innovation,

\begin{equation}
  \B{\nu}_{i,t} = \B{x}_{i,t} - \hat{\B{\mu}}_i - \hat{W}_i\,\hat{\B{z}}_{t\mid t-1}
\end{equation}

Under healthy conditions, (and the assumption of a well-specified model) these residuals are zero-mean, Gaussian, and i.i.d. Thus, the onset of damage can be inferred as a departure from their healthy distribution. To obtain estimates of the normal condition, a Gaussian is fitted to the training-window innovations for each structure,

\begin{equation}
  \hat{\B{\mu}}_i^{\nu}  = \mathbb{E}\left[\B{\nu}_{i, t}\right],
  \quad
  \hat{\Sigma}_i^{\nu} = \mathbb{V}\left[\B{\nu}_{i, t}\right],
  \quad \forall{t\in\text{Training window}}
  \label{eqn:normal_condition}
\end{equation}

and score every time step by the Mahalanobis distance (MD) to this empirical reference \cite{worden2000damage},

\begin{equation}
  D^2_{i,t} = (\B{\nu}_{i,t}-\hat{\B{\mu}}_i^{\nu})\tran (\hat{\Sigma}_i^{\nu})^{-1}(\B{\nu}_{i,t}-\hat{\B{\mu}}_i^{\nu})
  \label{eqn:md}
\end{equation}

For healthy structures, the MD follows a chi-squared distribution with $M$ degrees of freedom, $D^2_{i,t}\sim\chi^2_M$\footnote{The chi-squared statistic is strictly-speaking, an assumption, if the covariance matrix $\hat{\Sigma}_i^{\nu}$ is actually estimated from training data. However, it is a good approximation and the majority of outlier analysis studies in SHM make this simplification.}. A detection threshold $\tau$ is therefore obtained at a tolerable false-positive rate $\alpha$ as the corresponding upper quantile, $\tau = \chi^2_M(1-\alpha)$, the value exceeded by only a fraction of $\alpha$ healthy observations. Here, and throughout, $\chi^2_M(\cdot)$ denotes the quantile function (inverse CDF) of the chi-squared distribution with $M$ degrees of freedom.

\paragraph{Robust observation gating} A complication of the proposed approach in the population-based setting is that damage has the potential to bias the EOV estimation.  An anomalous observation, if assimilated by the Kalman filter, is partially absorbed into the shared latent $\B{z}_t$, potentially masking future  damage and contaminating the estimate for every other structure. To address this, an online gating strategy that down-weights highly anomalous observations is applied, in the spirit of robust Kalman filtering approaches that guard the state estimate against outlying measurements \cite{ting2007kalman, agamennoni2011outlier}.

At each test step and for each structure, a predictive gating distance $G^2_{i,t}$, is computed from the scaled Kalman innovations and predictive covariance,

\begin{equation}
  G^2_{i,t} = \frac{1}{d_i}\B{\nu}_{i,t}\tran S_{i,t}^{-1}\B{\nu}_{i,t}   
  \label{eqn:gate}
\end{equation}

with, 

\begin{equation}
  d_i = \frac{1}{M\,T_\text{train}}\sum_{t\in T_\text{train}}
    \B{\nu}_{i,t}\tran S_{i,t}^{-1}\,\B{\nu}_{i,t},
    \label{eqn:disp}
\end{equation}

being the average per-structure dispersion, estimated on the training window. This normalised-innovation-squared statistic is the same quantity used to form validation gates in target tracking \cite{barshalom1988tracking} and to test the whiteness of the innovation sequence for fault detection in dynamic systems \cite{mehra1971innovations}. Observations that exceed the threshold $G^2_{i,t}>\chi^2_M(1-\alpha_\text{gate})$ are reasoned to be sufficiently outlying that their inclusion is at risk of biasing the latent EOV removal and are thus excluded. Note that the gating procedure is applied per-observation. In this way, it is possible to exclude an outlier but include an observation afterwards. This is in contrast to some damage detection schemes that exclude all observations after an outlier is observed.

It is noteworthy that the gating procedure allows one to be much more aggressive with the threshold selection than might otherwise be possible during damage detection. The objective is not to classify observations as damaged, only to exclude samples \emph{suspected} of being damaged such that they do not bias the estimation. It is therefore possible to set the value of $\alpha_\text{gate}$ to a very permissive false-positive rate (FPR) without necessarily incurring a corresponding cost of many false activations in practice.

Finally, the Laplace posterior is propagated at each step to turn the point estimate into an uncertainty-aware decision signal. Drawing samples $\theta^{(s)}\sim\mathcal{N}(\hat\theta,\Sigma)$ and recomputing $D^2$ for each, the decision signal is thus the posterior exceedance probability, given by,

\begin{equation}
   \Pr\big(D^2_{i,t} > \tau\big) \approx \frac{1}{S}\sum_{s=1}^{S} \mathbb{I}\big[D^{2,(s)}_{i,t} > \tau\big]
   \label{eqn:exceed}
\end{equation}

for $S$ samples, where $\tau=\chi^2_M(1-\alpha)$ for some target FPR $\alpha$, and $\mathbb{I}[\cdot]$ is an indicator function. Values near zero indicate a confidently healthy structure, and near unity a confident detection. An overview of the proposed approach is given in Algorithm \ref{alg:gleov}.

\begin{algorithm}[h]
\caption{GLEOV: simultaneous EOV identification and damage detection}
\label{alg:gleov}
\begin{algorithmic}[1]
\Require features $\B{x}_{i,t}$, commissioning masks, lengthscale $\ell$, period $\Delta_t$,
         priors $p(\theta)$, $\alpha_\text{gate},\alpha$

\Statex \textbf{Identification} \Comment{training window}
\State form the Mat\'{e}rn SSM $(A_d,Q_d,P_\infty)$ from $(\ell,\Delta_t)$ \Comment{Eq.~\ref{eqn:ssm}}
\State $\hat{\theta}\gets$ MAP estimate \Comment{Eq.~\ref{eqn:map}, L-BFGS-B \cite{byrd1995lbfgsb, zhu1997lbfgsb}}
\State $\hat{\Sigma}\gets$ Laplace covariance; draw $\theta^{(s)}\sim\mathcal{N}(\hat\theta,\hat\Sigma)$, $s=1,\dots,S$ \Comment{Eq.~\ref{eqn:laplace}}
\State run KF at $\hat\theta$ over training; fit normal condition $(\hat{\B{\mu}}_i^{\nu},\hat{\Sigma}_i^{\nu})$ and dispersion $d_i$ \Comment{Eqs.~\ref{eqn:normal_condition},~\ref{eqn:disp}}

\Statex \textbf{Robust observation gating} %\Comment{causal forward pass at $\hat\theta$}
\For{$t=1,\dots,T$}
  \State KF predict; form innovation $\B{\nu}_{i,t}$ and predictive covariance $S_{i,t}$ \Comment{Eqs.~\ref{eqn:innova},~\ref{eqn:St}}
  \If{$t>T_\text{train}$ \textbf{and} $G^2_{i,t}>\chi^2_M(1-\alpha_\text{gate})$} \Comment{Eq.~\ref{eqn:gate}}
     \State mark $x_{i,t}$ as gated and exclude from observation model
  \EndIf
  \State KF update using non-gated observations
\EndFor

\Statex \textbf{Damage detection}
\For{$\theta'\in\{\hat\theta,\theta^{(1)},\dots,\theta^{(S)}\}$}
  \State run KF without gated observations; score $D^2_{i,t}(\theta')$ \Comment{Eq.~\ref{eqn:md}}
\EndFor
\State compute exceedance probability $p_{i,t}$ \Comment{Eq.~\ref{eqn:exceed}}
\State \Return MAP score $D^2_{i,t}(\hat\theta)$ and exceedance probability $p_{i,t}$
\end{algorithmic}
\end{algorithm}

\section{Case studies}

In order to demonstrate the effectiveness of the proposed GLEOV approach, two case-study examples of detection and removal of EOVs in SHM datasets are presented.

\subsection{Single structure}

Before demonstrating the proposed approach on a population of structures, it is important to first establish that the method is effective in identification and removal of EOV signals in the case of a single structure ($N=1$).  The first case study is therefore a laboratory-scale benchmark SHM dataset. The dataset under consideration comes from a Lamb-wave inspection investigation of a composite panel subject to temperature variations in an environmental chamber. The data were collected as part of the Brite-Euram project DAMASCOS (BE97 4213) \cite{cross2012features}. The dataset consists of Lamb-wave recordings collected every minute, comprising three distinct phases.

\begin{itemize}
    \item Phase I: For the first 1355 minutes, the chamber temperature was held at approximately 25°C.
    \item Phase II: For the next 1128 minutes, the chamber temperature was varied between 10°C and 30°C with a cycle time of approximately 370 minutes.
    \item Phase III: At minute 2483, the chamber was opened and damage was introduced by drilling a 10mm hole in the centre of the plate. Data collection continued under the same 10°C--30°C cycling; the present analysis uses the series up to minute 3700, giving 1217 minutes of post-damage monitoring.
\end{itemize}

\begin{figure}
  \centering
  \includegraphics[width=\columnwidth]{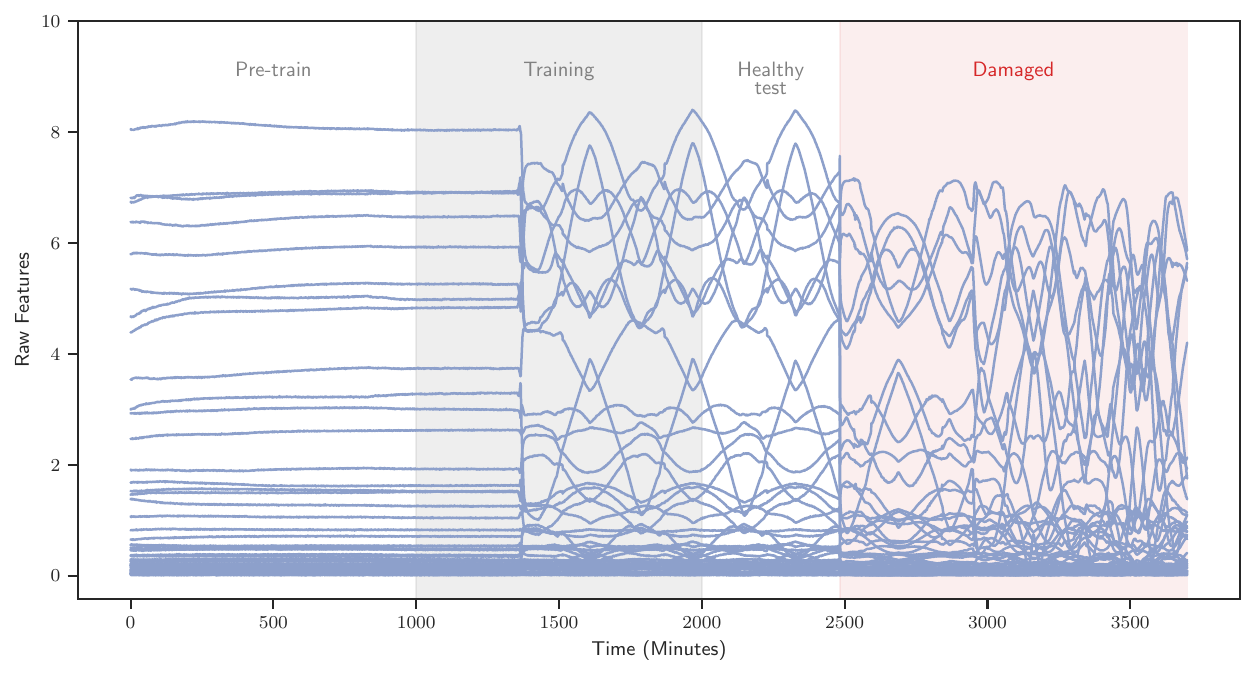}
  \caption{Raw features for the DAMASCOS dataset.}
  \label{fig:damascos_raw}
\end{figure}

The features under investigation are the magnitudes of 50 spectral lines in the vicinity of the peak of the frequency spectrum of the received Lamb-wave signal, these features are plotted in Figure \ref{fig:damascos_raw}. For a more complete description of the dataset and experimental setup, the interested reader is directed to \cite{cross2012features}. A training set (comprised of data from minutes 1000 to 2000) was extracted from the overall dataset. In order to capture both the stationary and non-stationary behaviour of the features, the training set comprises examples from both phases I and II. 

Because this dataset comprises only a single structure, an invariance exists between the two variance terms in equation~\eqref{eq:obs2}. As such, the two terms cannot be separately identified and so $\B\rho$ is removed from the model (i.e.\ $\tau_T=0$).

\begin{figure}
  \centering
  \includegraphics[width=\columnwidth]{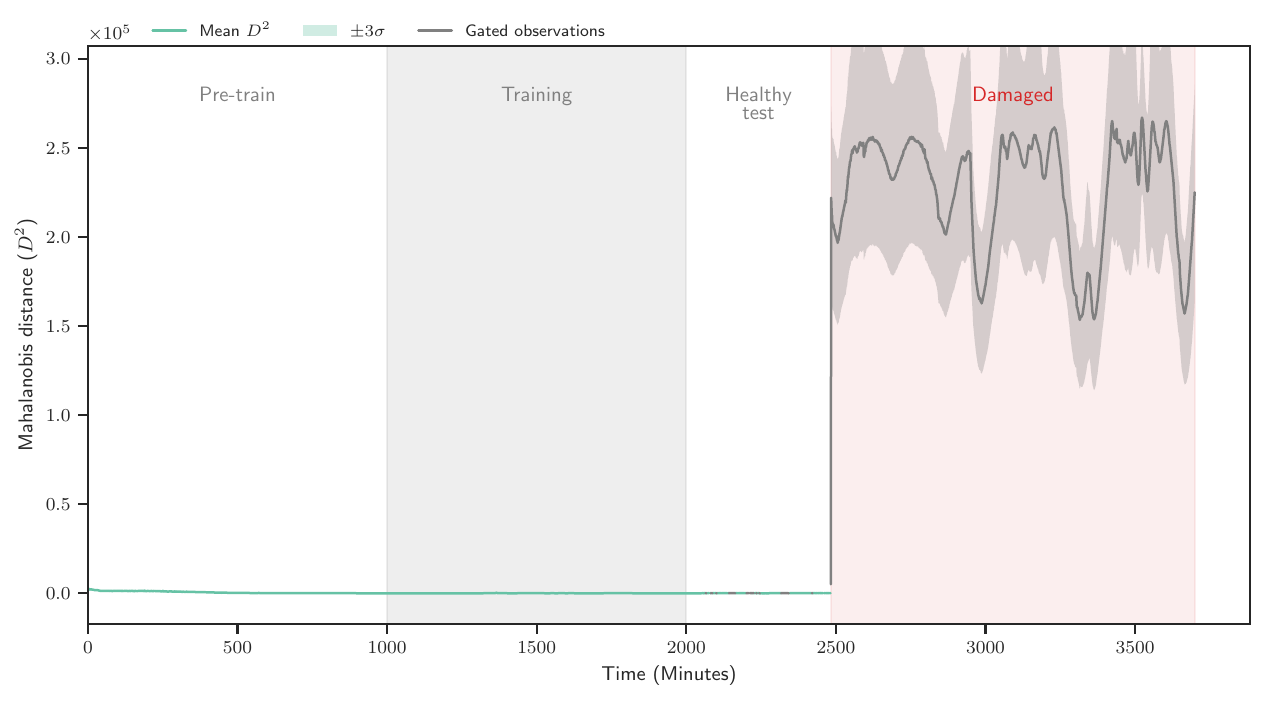}
  \caption{Mahalanobis $D^2$ detection trace (posterior samples) for the DAMASCOS dataset. Also shown (grey line) are the gated observations, for which $G^2_{i,t}>\chi^2_M(1-\alpha_\text{gate})$ that are excluded during prediction.}
  \label{fig:damascos_d2}
\end{figure}

The GLEOV algorithm was applied to this dataset as per Algorithm \ref{alg:gleov}, with prior distributions and parameters as in Table \ref{tbl:priors}. Here, the sampling period is $\Delta_t = 1$ minute, so the fixed lengthscale $\ell=100$ corresponds to $100$ minutes, roughly a quarter of the ${\sim}370$-minute thermal cycle. During the inference, $S=200$ Laplace posterior samples are drawn. The posterior predictive MD values ($D^2_t$) are plotted in Figure \ref{fig:damascos_d2}. The samples from the distribution are extremely sensitive to the presence of damage (AUC-ROC $=1.00$ over the test window), while remaining insensitive to the variation in temperature. Also visible in the figure are the gated observation points. It is clear that the gating has successfully removed all observations corresponding to damage as well as some of the more outlying points in the healthy testing regime. It is interesting that although several healthy observations are gated (22.6\%) this does not come at any cost to damage sensitivity or EOV reconstruction. 

\begin{figure}
  \centering
  \includegraphics[width=\columnwidth]{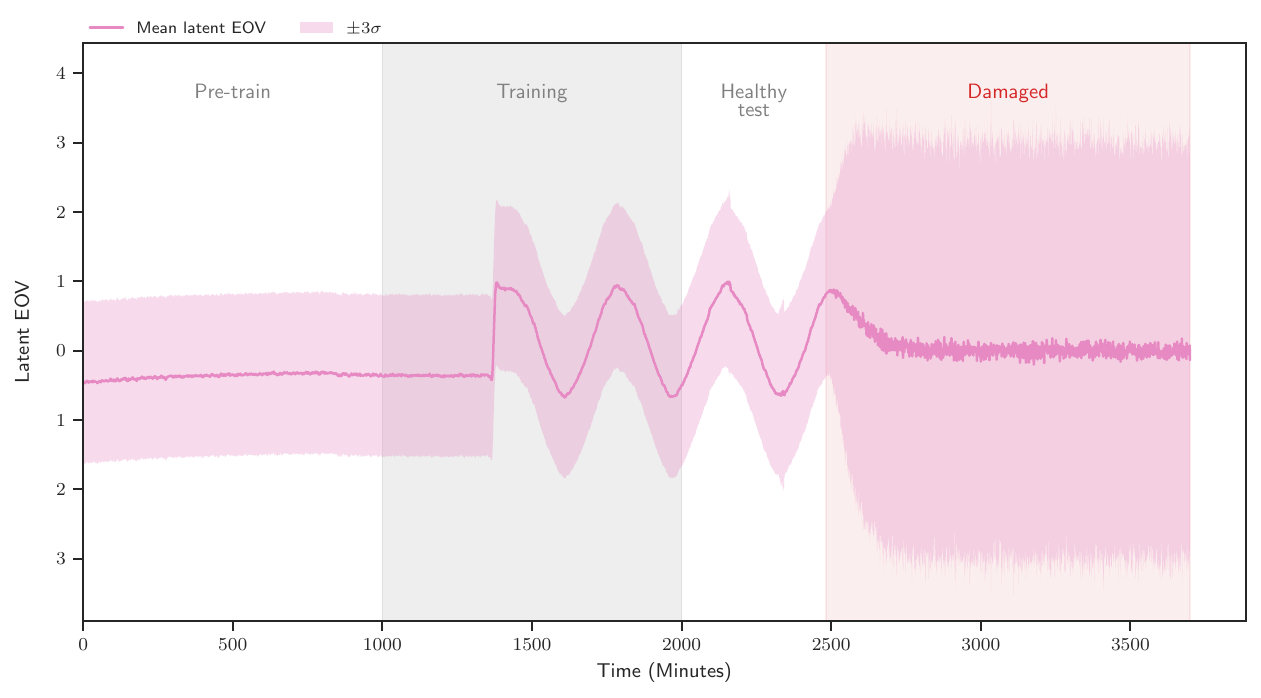}
  \caption{Inferred latent EOV signal $\hat{z}_t$ (posterior samples) over the full monitoring series, note that although the ground truth temperature is unavailable, good agreement is seen with the qualitative description of the test campaign in \cite{cross2012features}.}
  \label{fig:damascos_eov}
\end{figure}

Figure \ref{fig:damascos_eov} depicts the posterior over the inferred latent $\hat{z}_t$. It is remarkable that, despite never observing any temperature measurements, the recovered signal shows strong agreement with the qualitative thermal history of the campaign\footnote{The EOV is recovered up to an affine invariance because of the form of the data generating process. This invariance explains the apparently inverted EOV compared to the features in Figure \ref{fig:damascos_raw}.}. In the damage regime, all observations are excluded from the Kalman filter by the gating process and so in the absence of any information, the EOV returns correctly to the GP prior. In the context of a single structure, this behaviour is inevitable. However, in the context of a population of structures, the latent EOV can still be informed by healthy structures, as shown below. 

\subsection{Population of structures}
\label{sec:owt}

To demonstrate the effectiveness of the proposed approach on a population of structures, a simulated dataset corresponding to an offshore wind turbine farm subject to an unmeasured temperature EOV is introduced. This dataset has been deliberately designed to offer a robust modelling challenge in that the damage and EOV are highly confounded and that the EOV signal dominates the variance of the features. 

\paragraph{Simulated offshore wind turbine farm} 

The simulated population comprises $N=9$ nominally-identical turbines, each monitored via $M=3$ natural frequencies over $T=730$ daily samples, split into a $365$-day healthy training period and a $365$-day test period with a sampling period of $\Delta_t=1$ day. A single environmental driver (temperature), shared across the farm, is considered. It is generated as a Matérn-3/2 Gaussian process in time with a $60$-day lengthscale about a $9^\circ\text{C}$ mean. At a sampling rate of $\Delta_t=1$ day, daily temperature variations (e.g.\ day-night cycles) are considered to be averaged over and are not considered here. The temperature field is treated as common to the whole farm, with local Gaussian perturbations at each turbine. For features, the natural frequencies of the turbine's monopile are simulated; each turbine's features are set to depend affinely on its local temperature,

\begin{equation}
  \B{x}_{i,t} = \B{\mu}_i + \B{\phi}_i (\B{z}_t + \B{\rho}_{i,t}) + \B{\varepsilon}_{i,t}
\end{equation}

with,

\begin{equation}
  \B{\varepsilon}_{i,t}  \sim \mathcal{N}(\B{0},\; \sigma_\varepsilon^2 \B{I}_M),
  \quad
  \B{\rho}_{i,t} \sim \mathcal{N}(\B{0},\; \sigma_\rho^2 \B{I}_K)
\end{equation}

The per-turbine baseline frequencies $\B{\mu}_i$ and temperature sensitivities $\B{\phi}_i$ are sampled heterogeneously across the population to account for inter-structure variance. Damage is introduced asynchronously to five of the nine turbines, with onsets staggered across the test window days $485-665$. In order to ensure that this dataset represents a stiff challenge to SHM algorithms, damage is introduced primarily along the per-turbine thermal loading vector $\B{\phi}_i$. 

\begin{equation}
  \Delta \B{x}_{i,t}
  = \beta_i \left( \B{\phi}_i + 0.1 \,\|\B{\phi}_i\|\, \B{g}_\perp \right)
  \label{eq:damage}
\end{equation}

with $\B{g}_\perp$ a unit vector orthogonal to $\B{\phi}_i$ and where $\beta_i$ is a damage severity parameter. Thus, the damage signal is highly confounded with the EOV, with only a small component that acts independently. The motivation for this choice of damage model (other than it being challenging from an identification point of view) is that under a proportional stiffness perturbation $\mathcal{K}\to \mathcal{K}(1+\beta)$ with mass unchanged (caused by either temperature variation or stiffness-reducing damage, e.g.\ scour) the natural frequencies shift as $f_n \to f_n\sqrt{1+\beta} \approx f_n(1 + \beta/2)$ to first order. The induced feature-space displacement therefore lies along the same direction as the thermal loading, independent of the severity. This suggests that when EOVs and damage both cause proportional stiffness changes, the resulting changes to the natural-frequency features can lie in the same subspace. 

\begin{figure}
  \centering
  \includegraphics[width=\columnwidth]{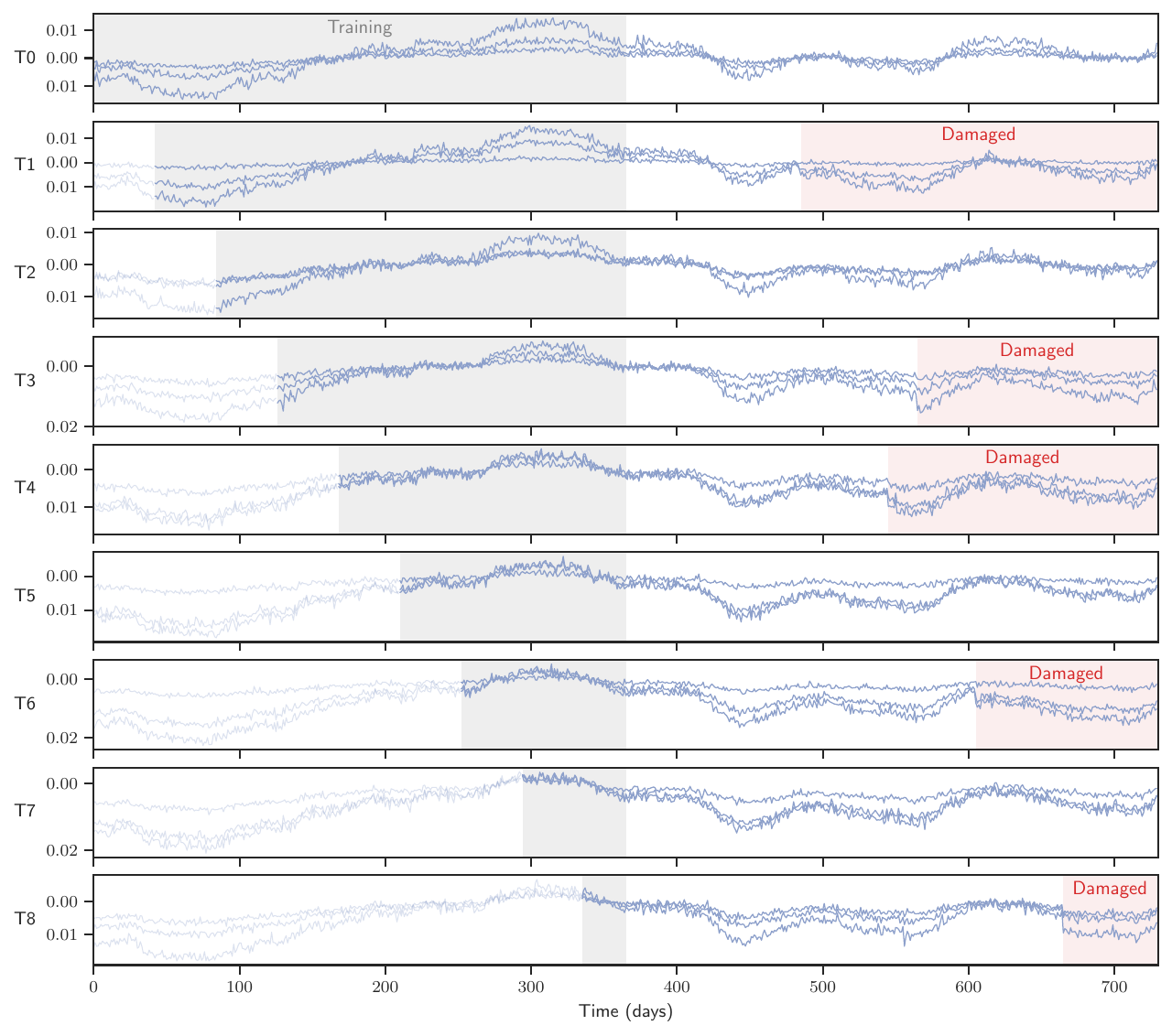}
  \caption{Raw natural frequency features (scaled to zero-mean for plotting) for the simulated OWT farm.}
  \label{fig:latent_raw}
\end{figure}

In order to demonstrate the benefits of the GLEOV approach in a population-based context, the dataset implements a synthetic staggered deployment. For each turbine in the population, the number of data available for training is spaced linearly from the full $365$ down to $30$ across the nine turbines. The simulated features, training windows and damage regions are depicted in Figure \ref{fig:latent_raw}. All parameter values pertaining to the creation of the synthetic OWT farm are collected in Table \ref{tbl:owt_params}. It is clear from the figure that this dataset represents a robust modelling challenge. The onset of damage is barely perceptible and is dwarfed by the variation caused by the EOV. Furthermore, several turbines have remarkably few data available for training, particularly T8 which sees only a 30-day window.

\begin{table}
  \centering
  \caption{Parameters of the simulated nine-turbine offshore wind farm.}
  \label{tbl:owt_params}
  \begin{tabular}{l|ll}
    \hline
    Parameter & Symbol & Value \\\\
    \hline
    \multicolumn{3}{l}{\emph{Dimensions}} \\
    \hline
    \quad Turbines                    & $N$ & $9$ \\
    \quad Frequencies per turbine     & $M$ & $3$ \\
    \quad Duration (train\,/\,test)   & $T$ & $730$ ($365$\,/\,$365$) days \\
    \quad Training days per turbine   &       & $365 \to 30$ (linearly spaced) \\
    \hline
    \multicolumn{3}{l}{\emph{Temperature EOV}} \\
    \hline
    \quad Mean temperature            & $T_\text{base}$ & $9.0^{\circ}$C \\
    \quad GP amplitude standard deviation        &                 & $4.0^{\circ}$C \\
    \quad GP lengthscale              & $\ell$          & $60$ days \\
    \quad Per turbine noise standard deviation    & $\sigma_\rho$          & $0.5^{\circ}$C \\
    \hline
    \multicolumn{3}{l}{\emph{Feature model}} \\
    \hline
    \quad Baseline frequencies (per turbine)    & $\B{\mu}$   &  \\
    \qquad mean    &    & $[0.30,\,0.65,\,1.05]$ Hz \\
    \qquad standard deviation           &         & $[5,\,10,\,15]\times10^{-3}$ Hz \\
    
    \quad Temperature sensitivity (per turbine) & $\B{\phi}$  &  \\
    \qquad mean          &       & $[-5,\,-10,\,-15]\times10^{-4}$ Hz$^{\circ}$C$^{-1}$ \\
    \qquad standard deviation              &       & $[1,\,2,\,3]\times10^{-4}$ Hz$^{\circ}$C$^{-1}$ \\
    \quad Frequency noise standard deviation         & $\sigma_\varepsilon$        & $5\times10^{-4}$ Hz \\
    \hline
    \multicolumn{3}{l}{\emph{Damage model}} \\
    \hline
    \quad Damaged turbines            &      & $5$ of $9$ \\
    \quad Severity (per damaged turbine)                   & $\beta_i$ & $2.5$--$3.0$ \\
    \quad Off-axis fraction           &    & $0.10$ \\
    \quad Onset range                 & $t_i^*$    & days $485$--$665$ \\
    \hline
  \end{tabular}
\end{table}

\begin{figure}
  \centering
  \includegraphics[width=\columnwidth]{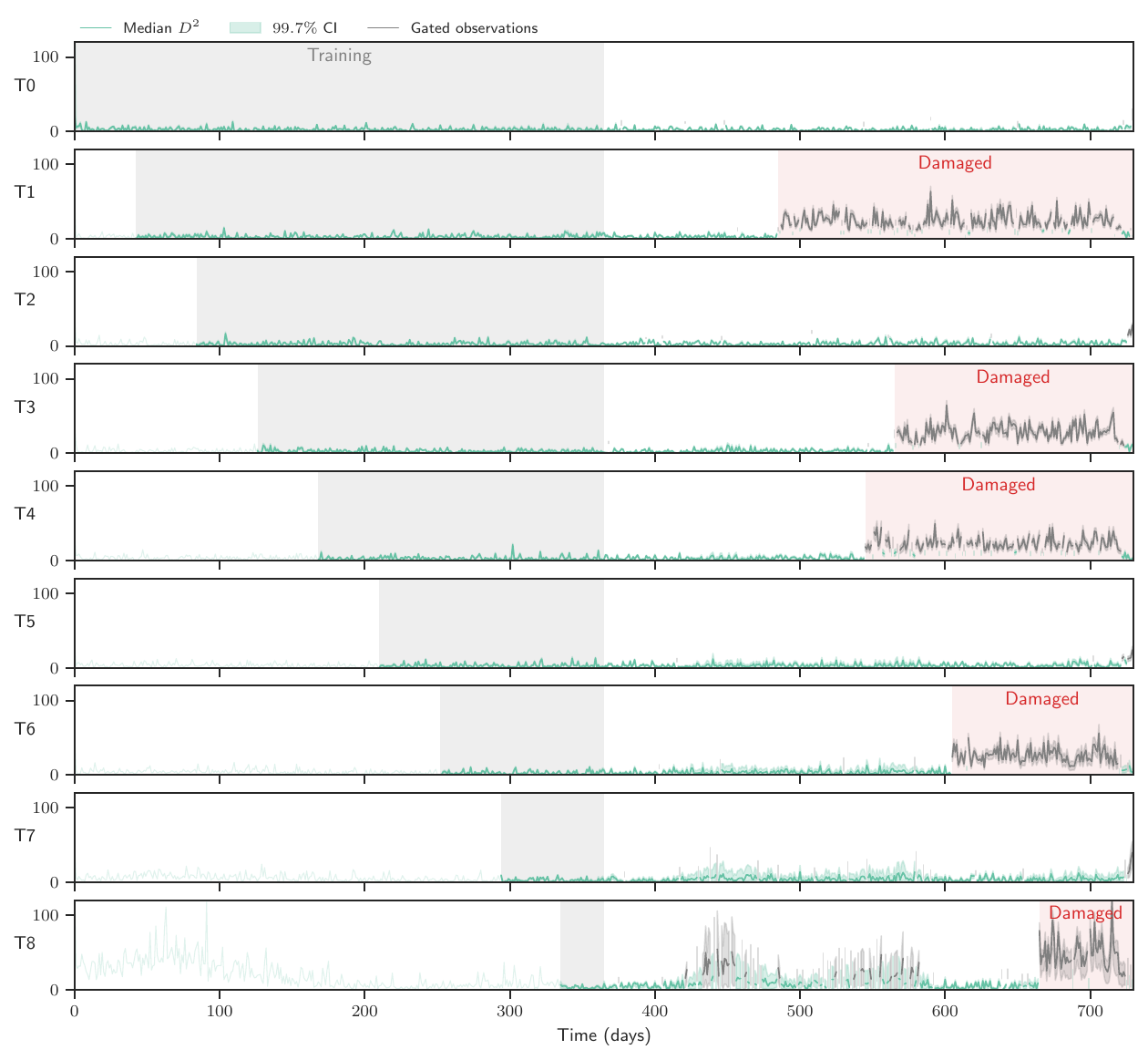}
  \caption{GLEOV posterior distribution over the Mahalanobis distances $D^2_{i,t}$, for the simulated OWT farm dataset. Also shown (grey line) are the gated observations, for which $G^2_{i,t}>\chi^2_M(1-\alpha_\text{gate})$ that are excluded during prediction.}
  \label{fig:latent_D2}
\end{figure}

With the dataset established, the proposed GLEOV approach is applied as described above; the sampling period is $\Delta_t=1$ day, so the fixed lengthscale $\ell=100$ corresponds to $100$ days. During the inference, $S=500$ Laplace posterior samples are drawn. In Figure \ref{fig:latent_D2}, the posterior predictive MD $D^2_{i,t}$ are plotted for all turbines in the population. As can be seen in the figure, the sensitivity to damage is excellent. Furthermore, the gated observations are plotted (coloured in grey). Here, it can be seen that almost all observations from damaged regimes are excluded. Although some false positive readings are visible in T8, it is encouraging that these readings are also excluded by the gating mechanism so that these observations do not bias the estimation of the latent EOV.  

\begin{figure}
  \centering
  \includegraphics[width=\columnwidth]{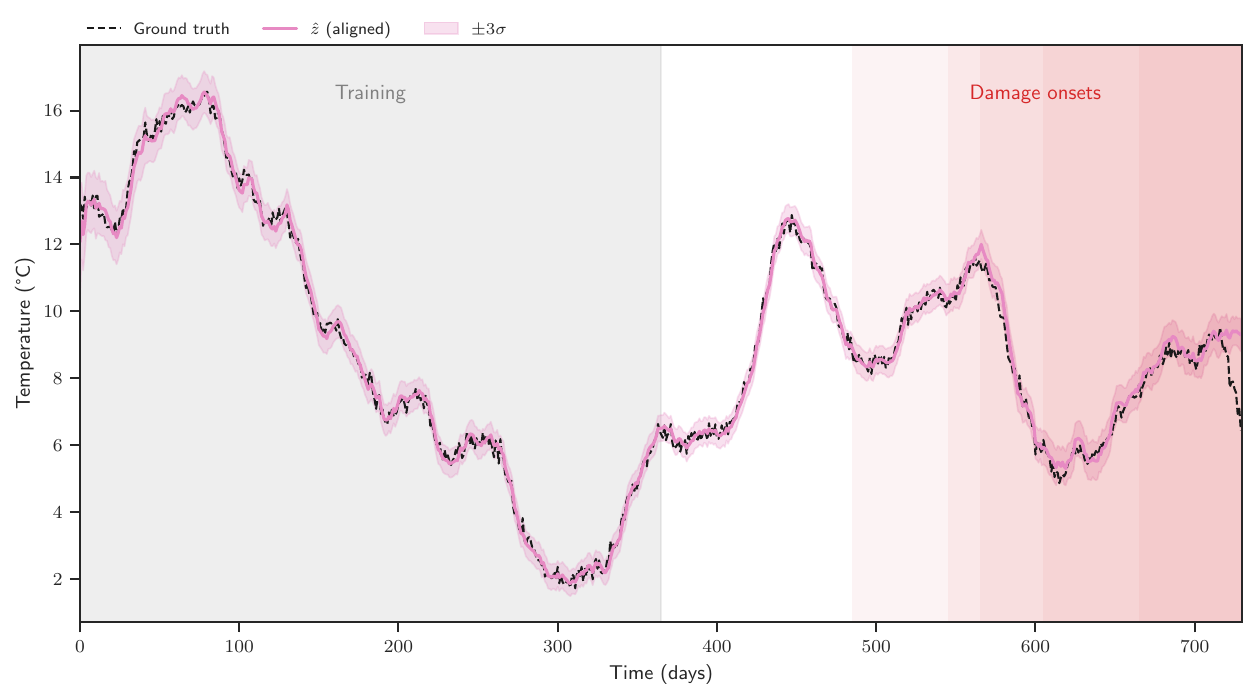}
  \caption{GLEOV posterior predictions of the latent temperature EOV $\hat{z}_t$ for the simulated OWT farm dataset, for plotting these predictions are aligned affinely to the ground truth signal by least squares on the training data. The darkening red regions indicate damage onset events in the population with darker shading corresponding to a greater damaged proportion of the population.}
  \label{fig:latent_eov}
\end{figure}

The latent EOV signal is plotted in Figure \ref{fig:latent_eov}. Because this is a simulated dataset, the ground truth is available (although there is an affine invariance). In the figure, the latent EOV is aligned to the ground truth by a least-squares fit to the affine transformation on the training window. As can be seen in the figure, the posterior prediction of the EOV is excellent. As expected, uncertainty shrinks over the course of the training window as data from more turbines become available. At the onset of damage, observations from damaged turbines are excluded, and the tracking remains strong. At the very end of the test window (with damage in over half of the population), performance begins to degrade and the tracking diverges slightly from the ground truth.  

\begin{figure}
  \centering
  \includegraphics[width=\columnwidth]{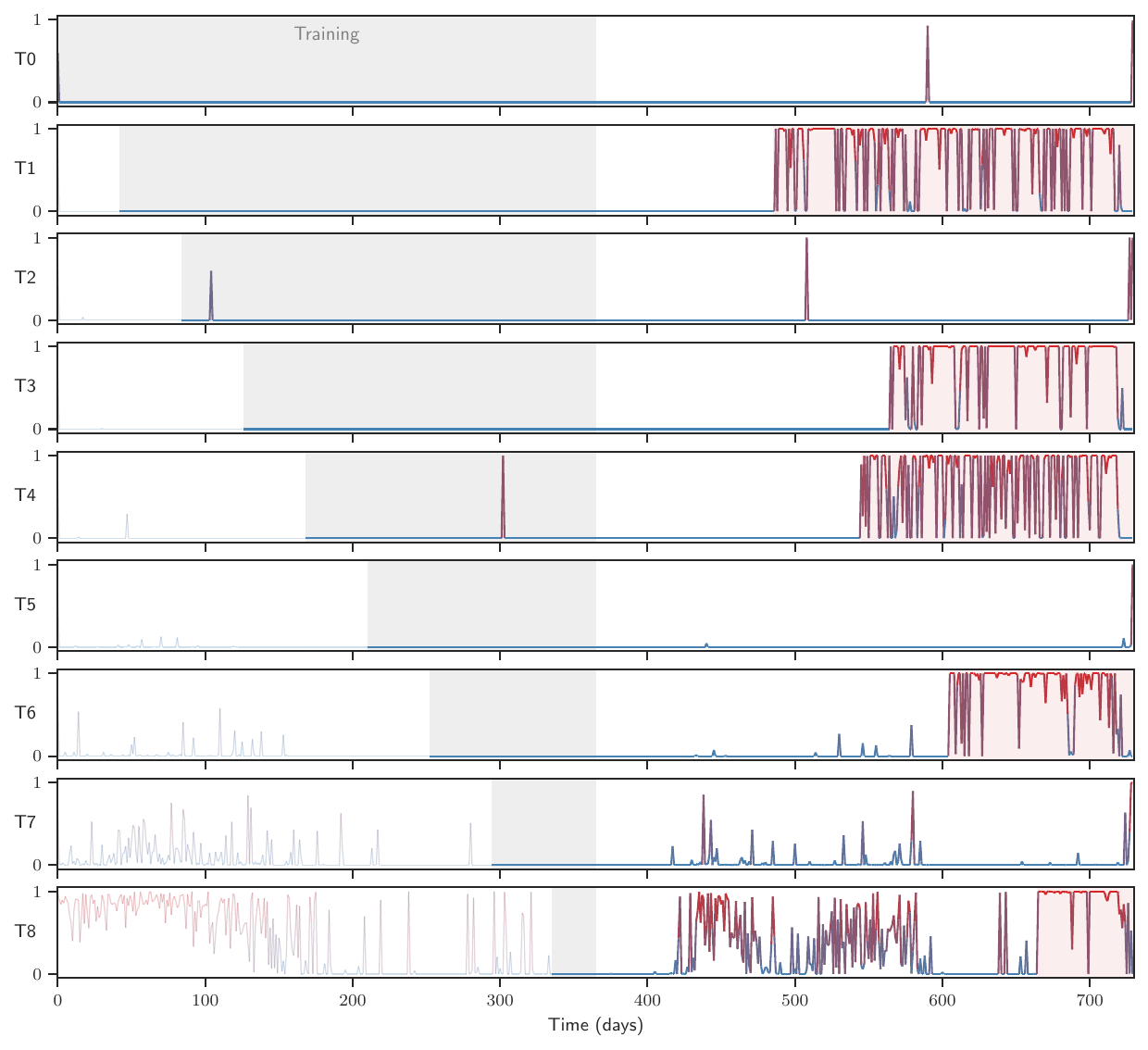}
  \caption{GLEOV exceedance probability $p_{i,t}$ for each turbine in the OWT farm dataset at a threshold level targeting 0.001 FPR on the training window.}
  \label{fig:latent_p}
\end{figure}

In Figure \ref{fig:latent_p}, the exceedance probability is plotted for each turbine in the population, with the threshold set to a target FPR of $\alpha=0.001$ on the training window. As can be seen in the figure, the exceedance probability is high in the damage regimes and remains close to 0 at almost all other times for all turbines. The exception to this trend is T8, the most data-poor member of the population, commissioned with only $30$ days of healthy training data. This is not an unexpected result, for T8, the EOV loading direction $W_8$ is only weakly identified, and the corresponding posterior over the $D^2$ is broad (as can be seen in Figure \ref{fig:latent_D2}). The exceedance probability inherits this uncertainty. Although the exceedance probability is raised during the healthy portion of the testing window for T8, it remains lower than during the damaged region, indicating sensitivity to damage.     

\paragraph{Baseline comparison} 

GLEOV is compared here against five baselines, each scored by the same Mahalanobis distance statistic (equation \eqref{eqn:md}) to provide a fair comparison between methods. The first, \emph{raw features}, performs no EOV removal and detects damage directly in the $M$-dimensional feature space. The next two are variants of projection-based \emph{minor-component analysis} (MCA) \cite{yan2005partI, yan2005partII}, the canonical variance-dominance approach. In MCA, the top-$K$ principal directions of the healthy training features (assumed to span the dominant environmental variation) are discarded by singular value decomposition, and detection is carried out in the residual $(M-K)$-dimensional minor-component subspace. The two MCA variants differ only in how this subspace is estimated. \emph{Per-structure} MCA estimates it independently for each turbine, whereas \emph{pooled} MCA estimates a single shared subspace jointly from the whole population.

The final two are \emph{cointegration} baselines, the other canonical implicit EOV-removal paradigm \cite{cross2011cointegration, cross2012cointegration}. Here the features are assumed to share a common non-stationary environmental trend, and a stationary linear combination of them is sought that cancels this trend. Damage is then inferred by a loss of stationarity of that combination. The cointegrating vector is estimated by the Johansen procedure \cite{johansen1991estimation}, following the approach of \cite{cross2011cointegration}. Detection is carried out on the resulting one-dimensional cointegration residual. As for MCA, a \emph{per-structure} variant estimates the cointegrating vector independently per turbine, while a \emph{pooled} variant estimates a single shared vector across the population.

Rather than present results at only a single threshold level, it is of interest to consider the performance of the GLEOV method and baselines across threshold levels. A convenient way to achieve this is by computing the receiver operating characteristic (ROC) curve \cite{murphy2023probml2}. The ROC traces the true-positive rate (TPR) against the false-positive rate (FPR) as the detection threshold is swept across its full range. A convenient single scalar summary value is available as the area under the ROC curve (AUC-ROC), which is equivalent to the probability that a randomly-chosen damaged observation is scored higher than a randomly-chosen healthy one \cite{murphy2023probml2}, so that $\text{AUC}=1$ denotes perfect separation and $\text{AUC}=0.5$ implies chance.

\begin{figure}
  \centering
  \includegraphics[width=0.6\columnwidth]{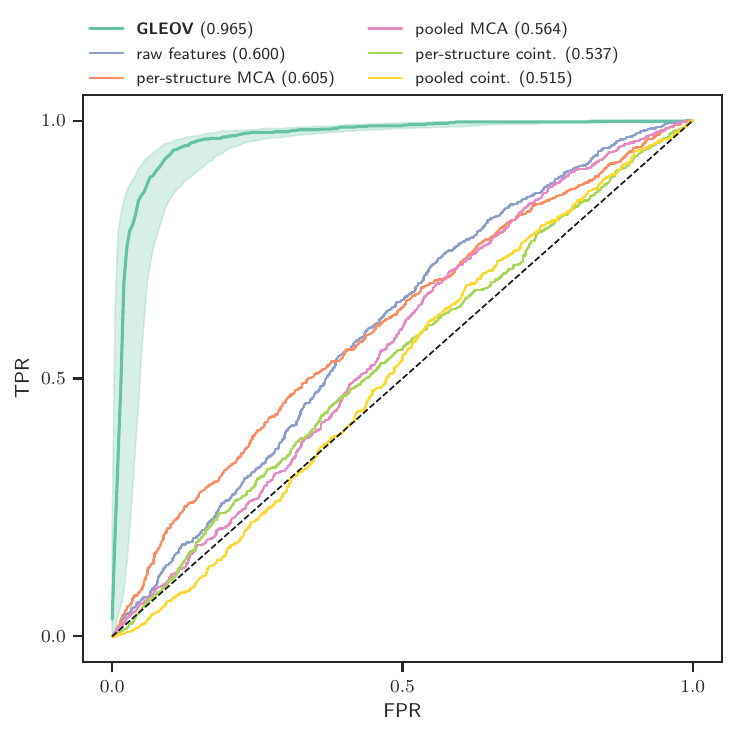}
  \caption{Pooled ROC curves over all OWT test observations for GLEOV and the five baselines.}
  \label{fig:roc}
\end{figure}

Figure \ref{fig:roc} shows the pooled ROC over all test observations for GLEOV and the five baselines. GLEOV attains an AUC of $0.965$ in the OWT test and lies above every baseline at every operating point. The baselines remain close to the diagonal (per-structure MCA $0.605$, raw features $0.600$, pooled MCA $0.564$, per-structure cointegration $0.537$ and pooled cointegration $0.515$), barely exceeding chance. This gap is a direct consequence of the adversarial damage design of equation \eqref{eq:damage}. Because the damage acts predominantly along the EOV direction, the baselines that remove the environmental subspace (either by discarding its high-variance directions (MCA) or by cancelling the shared environmental trend with a stationary combination (cointegration)) remove the damage signature along with it and retain only the small off-axis residual. By identifying the EOV via its slowness rather than its variance, GLEOV separates damage from the environment \emph{in time} and recovers the collinear component that the baselines remove.

The shaded region in Figure~\ref{fig:roc} quantifies the effect of parameter uncertainty. Each Laplace posterior sample yields its own ROC, and the band spans their $99.7\%$ central interval, with the ROC curve corresponding to the MAP curve shown as the central line. The band is narrow, indicating that the posterior uncertainty in $\theta$ has little bearing on the relative performance of the GLEOV and baseline methods.

\begin{figure}
  \centering
  \includegraphics[width=\columnwidth]{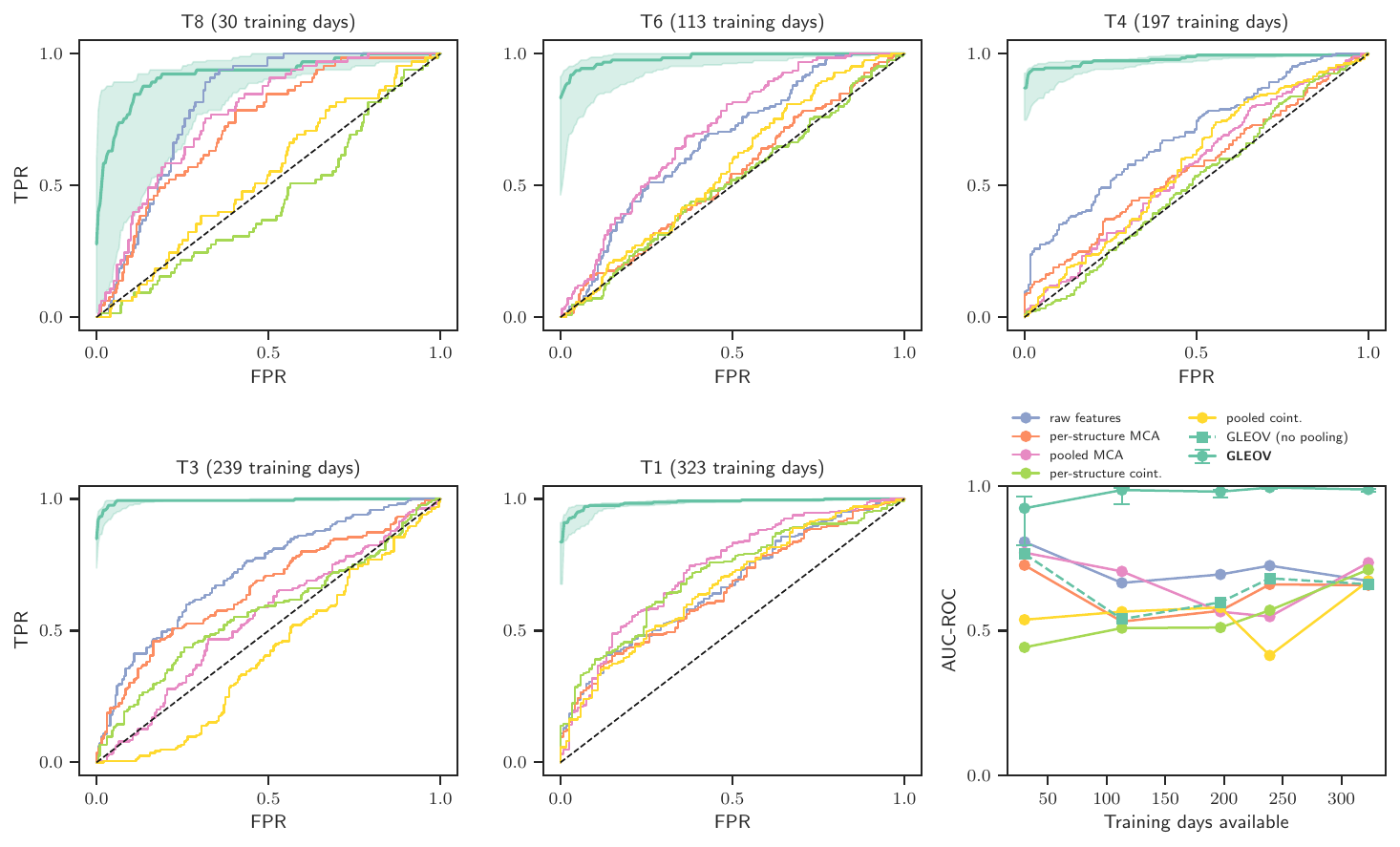}
  \caption{Per-turbine ROC curves for each of the five damaged turbines, comparing GLEOV against the raw-feature, MCA and cointegration baselines. The final panel plots AUC against training-set size for each method, together with the no-pooling ablation of GLEOV in which each turbine is fit independently by the GLEOV approach.}
  \label{fig:pbshm}
\end{figure}

In Figure \ref{fig:pbshm}, the ROC curves are reported for each of the damaged OWTs in the dataset independently. GLEOV (solid) attains per-turbine AUCs of $0.92$--$0.99$ in every panel, while the projection and cointegration baselines stay near the diagonal ($0.41$--$0.77$). The lower-right panel plots AUC against training-set size, with GLEOV high and approximately flat (dropping only to $0.92$ on the most data-poor turbine) and the projection baselines below it at every level. The same panel also reports an ablation study in which the GLEOV population sharing is switched off and each turbine is fit independently, with no shared consensus loading $W_0$ and no shared latent EOV (\emph{GLEOV, no pooling}). Stripped of the population-level data sharing, detection collapses to the per-structure MCA at every training-set size ($0.54$--$0.77$). 

This result is a clear motivation for the population-based approach proposed in this work; the shared EOV is anchored by the healthy members, while the data-poor loadings are additionally shrunk toward the consensus $W_0$. The most data-poor turbine, T8, with only $30$ days of data, still benefits measurably from the pooling (AUC $0.77 \to 0.92$).

\section{Discussion} 

This work has presented GLEOV, a Bayesian recursive-estimation framework for the simultaneous identification and removal of EOVs in population-based structural health monitoring. The unobserved EOV is represented by a latent Gaussian process in time, shared across the population and identified by its \emph{slowness} rather than its variance. Cast in linear-Gaussian state-space form, the GP is marginalised exactly by a Kalman filter at $\mathcal{O}(T)$ cost, while a hierarchical prior over the per-structure loadings lets data-rich structures lend statistical strength to data-poor ones. The approach has been demonstrated on two case studies: A single laboratory-scale structure subject to thermal variability, and a simulated population of offshore wind turbines with damage acting along the environmental direction under staggered deployment.

In the first case study, the framework was validated on real monitoring data from a single laboratory-scale structure subject to thermal variability. Without observing temperature, GLEOV recovered the latent environmental signal directly from the measured features, in agreement with the qualitative thermal history of the test campaign. The resulting EOV-removed residuals are highly damage sensitive ($\text{AUC}=1.00$). This study demonstrates identification and detection performance on real-world data, before the benefits of population sharing are examined in the second case study.

The second case study examined the population setting on a simulated nine-turbine farm, constructed deliberately so that damage acts along the same feature-space direction as the temperature EOV. Here, GLEOV recovers with uncertainty the damage signal that is missed by the projection baselines. The pooled ROC is above every baseline at every operating point ($\text{AUC}=0.965$) compared to the baseline results ($0.52$--$0.61$). This robustness to data scarcity is the central benefit of treating EOV removal within a population-based framework; the joint identification of shared latent EOVs means that the data-poor members are better able to remove the effect of the EOV by borrowing strength from other turbines with more observations. 

Modelling a shared latent EOV is, however, not without risk. An anomalous observation on one structure can be absorbed into the common EOV and bias the estimate for the whole population. The gating strategy proposed in this work guards against this by withholding observations inconsistent with the predicted environmental response from the latent update. In this way, damage cannot be accidentally interpreted as environmental variation, overcoming one of the main limitations of SFA in SHM. 

A further advantage of the proposed gating is that it can be operated far more permissively than the detection threshold. An operator would be free to gate observations liberally (accepting that some observations would be falsely excluded) without the corresponding cost of false positive activations (potentially very expensive if many superfluous inspections are triggered).  

The Bayesian framework adopted in this work lends further robustness to the proposed detection approach. Propagating the Laplace posterior gives access to an uncertainty-aware exceedance probability that correctly concentrates uncertainty on the structures with the lowest amount of data available.

Although the proposed approach is shown here to be highly promising, there remain some important limitations. The first, is that the Laplace approximation, being a Gaussian surrogate centred at the mode, can be inaccurate when the true posterior departs substantially from Gaussianity \cite{kuss2005assessing}. Future work by the authors will address this by employing more advanced Bayesian inference techniques (Hamiltonian Monte-Carlo, Stochastic variational inference). Another limitation is the choice of an isotropic noise structure for $\B\epsilon_{i,t}$ and $\B\rho_{i,t}$. If the EOVs or EOV weights vary on drastically different timescales this strong assumption may not be warranted and a more rich correlation structure would have to be considered. Also of interest are alternative methods for the MAP optimisation procedure used herein. Methods such as expectation maximisation \cite{dempster1977maximum, shumway1982approach} may offer performance benefits over the LBFGS-B approach. A final limitation is that a single latent EOV dimension ($K=1$) is used throughout both case studies, reflecting the single dominant EOV in each dataset. Where several environmental factors act concurrently (for example, temperature alongside wind or humidity), a higher-dimensional latent ($K>1$) would be required. The state-space formulation accommodates this directly by inflating the latent state, but its behaviour in the population setting remains to be explored and is left to future investigation.

In this work, the lengthscale parameter of the Gaussian process has been treated as a fixed hyperparameter rather than inferred. Although the hyperparameter sensitivity study in Appendix \ref{app:lengthscale} demonstrates broad robustness to this choice, incorporating $\ell$ as an inference parameter under a prior that enforces slowness nonetheless remains a promising avenue for future work.

In conclusion, unobserved environmental and operational variability remains a central obstacle to reliable SHM in a population-based context. By identifying the latent EOV through its temporal correlation, sharing it across a population, and robustly gating anomalous observations, GLEOV recovers the confounding trend rather than discarding variance, sharpening the separation between benign environmental change and genuine damage and offering a promising route towards more robust operation and maintenance decision-making.

\section*{Statements and declarations}

\subsection*{Ethical considerations}

Not applicable.

\subsection*{Consent to participate}

Not applicable.

\subsection*{Consent for publication}

Not applicable.

\subsection*{Declaration of conflicting interest}

Not applicable.

\subsection*{Funding}

The authors of this paper gratefully acknowledge the support of the UK Engineering and Physical Sciences Research Council (EPSRC) via grant reference EP/W005816/1 (ROSEHIPS). For the purpose of open access, the authors have applied a Creative Commons Attribution (CC BY) licence to any Author Accepted Manuscript version arising.

EJC, TJR and MRJ would like to acknowledge the support of Innovate UK through the OLLGA project grant 10040817.

MDC gratefully acknowledges the support of the Centre for Machine Intelligence (CMI) within the University of Sheffield.

MRJ gratefully acknowledges the support of the University of Sheffield through a Research Excellence Fellowship.

\subsection*{Data availability}

The experimental data analysed in the first case study were collected under the Brite-Euram project DAMASCOS (BE97 4213) and are not owned by the authors; a  description of the dataset and experimental campaign is given in \cite{cross2012features}. 

The code implementing the GLEOV model and all baseline methods will be made available upon acceptance of the paper.
\bibliography{eovs}

\appendix

\section{Lengthscale sensitivity}
\label{app:lengthscale}

\begin{figure}
  \centering
  \includegraphics[width=\columnwidth]{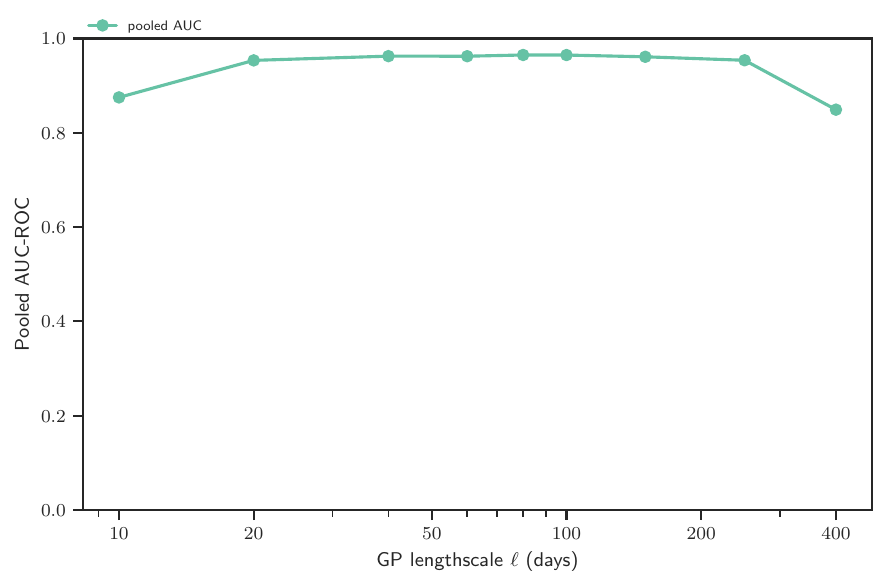}
  \caption{Pooled population AUC-ROC for the OWT case study as the GP lengthscale $\ell$ is swept, with all other hyperparameters and priors fixed as in the main text.}
  \label{fig:lengthscale}
\end{figure}

Throughout this work the Gaussian-process lengthscale $\ell$ is treated as a fixed hyperparameter rather than inferred, encoding the \emph{slowness} assumption that the latent EOV varies on longer timescales than damage, noise or the structural dynamics. To verify that the reported performance does not hinge on the precise value chosen, the population case study of Section~\ref{sec:owt} is re-run across a grid of lengthscales, with all other priors and hyperparameters held at the values of Table~\ref{tbl:priors}. At each $\ell$, the full pooled model is refit and the pooled population AUC-ROC over the test window is reported. The results are shown in Figure~\ref{fig:lengthscale}.

As can be seen in the figure, the AUC-ROC performance across the population remains largely insensitive to $\ell$ across a large range of choices of lengthscale, encompassing the range of values that might be selected to model annual seasonal variation. AUC-ROC remains at $\approx 0.96$ for lengthscales between $40$--$150$ days, comfortably bracketing both the $60$-day lengthscale of the simulated environmental field and the $\ell=100$ days value adopted in the main text. 

Damage identification performance degradation is only seen at the extremes, and even there it remains far above the best projection baseline ($0.61$, Section~\ref{sec:owt}). The fixed choice of $\ell$ is therefore not a significant limitation of the results in this work.

\end{document}